\documentclass[aps,PRB,twocolumn,superscriptaddress,preprintnumbers]{revtex4-2}
\usepackage[T1]{fontenc}
\usepackage{times}
\usepackage{graphicx,color}
\usepackage{amsfonts,amsmath,amssymb,amsbsy}

\begin{document}

\title{Higher-order bulk photovoltaic effects, quantum geometry and
application to $p$-wave magnets}
\author{Motohiko Ezawa}
\affiliation{Department of Applied Physics, The University of Tokyo, 7-3-1 Hongo, Tokyo
113-8656, Japan}

\begin{abstract}
The injection and shift currents are generalized to the $\ell $th-order
injection and shift currents for the longitudinal conductivities in the
two-band model, where $\ell $ is the power of the applied electric field. In
addition, the formulas for the higher-order injection current are expressed
in terms of the quantum metric and the higher-order shift current in terms
of the higher-order quantum connection. Then, they are applied to $p$-wave
magnets. It is shown that the injection and shift currents are zero. On the
other hand, the $\ell $th-order injection and shift currents with odd $\ell $
are nonzero when the direction of the N\'{e}el vector of the $p$-wave magnet
points to an in-plane direction.
\end{abstract}

\date{\today }
\maketitle

\section{Introduction}

Photovoltaic currents are generated under photo-irradiation\cite%
{Beli,Kraut,Baltz,Ave,Sipe,Frid}, which will be useful for solar cell
technologies. A $p$-$n$ junction presents a conventional way to generate
photocurrent. On the other hand, the bulk photovoltaic photocurrent
generation\ presents another way without using a junction\cite%
{Beli,Kraut,Baltz,Ave,Sipe,Frid}. The injection current\cite%
{Sipe,JuanNC,Juan,Ave,AhnX,AhnNP,WatanabeInject,Okumura,Dai,EzawaVolta} and
the shift current\cite%
{Young,Young2,Ave,Kraut,Baltz,Sipe,Juan,AhnX,MorimotoScAd,Kim,Barik,AhnNP,WatanabeInject,Dai,Yoshida,EzawaVolta}
are prominent, which are second-order bulk photovoltaic currents. Similarly,
third-order bulk photovoltaic currents are studied\cite%
{Sipe,Ave,MorimotoScAd,Jerk,JerkComment,JerkReply,Park,snap,AhnNP,Ma2}. The
jerk current which is the third order generalization of the injection
current is known to be dominant in the clean limit\cite%
{Jerk,JerkComment,JerkReply,snap,AhnNP}.

Quantum metric is a quantum geometric quantity\cite{Provost,Ma}, which is
the real part of the quantum geometric tensor. On the other hand, the Berry
curvature is the imaginary part of the quantum geometric tensor. The quantum
metric is observable by means of optical absorption\cite%
{AhnX,Holder,Bhalla,AhnNP,Souza,WChen2022,Onishi,Sousa,Ghosh,WChen2024,WChen2025,EzawaQG,Oh,JiangRev}
and\ the injection current \cite{AhnNP}.

The $p$-wave magnet is characterized by the $p$-wave spin splitting in the
energy spectrum\cite{pwave,Martin,H20,H20b,H22,Okumura,Kuda}. Its essence is
described by a two-band model\cite%
{Yamada,EzawaPwave,EzawaPNeel,Edel,EzawaVolta}. They were recently realized
experimentally in\cite{Yamada} Gd$_{3}$Ru$_{4}$Al$_{12}$ and in\cite{Comin}
NiI$_{2}$. However, mainly the combined effect with the Rashba interaction%
\cite{EzawaPwave,EzawaPNeel,Edel,EzawaVolta,Elliptic} or superconductivity%
\cite{Maeda,EzawaPwave,Fukaya} has been studied theoretically. It is
interesting to search for phenomena occurring without the Rashba interaction
or superconductivity in the $p$-wave magnet.

In this paper, we generalize the injection current and the shift current to
the $\ell $th-order photocurrents with an arbitrary $\ell $ for the
longitudinal conductivities in the two-band model. The $\ell $th-order
injection (shift) current dominates the $\ell $th-order shift (injection)
current in the clean (dirty) sample limit. Then, we express the higher-order
injection current in terms of the quantum metric and the higher-order shift
current in terms of the higher-order quantum connection. Then, we apply the
results to $p$-wave magnets in the presence of magnetic field. We first show
that the injection current and the shift current are exactly zero. Next, we
show that the\textsl{\ }odd-order injection and shift currents takes nonzero
value when the direction of the N\'{e}el vector of the $p$-wave magnet
points to an in-plane direction. Furthermore, we obtain their analytic form.

\section{Nonlinear conductivity}

The current density $J^{c}$ along the $c$ direction induced by the applied
electric field $E_{x}$ along the $x$ direction is expanded in a power series
of $E_{x}$ as%
\begin{equation}
J^{c}=\sum_{\ell =1}^{\infty }\sigma ^{c;x^{\ell }}E_{x}^{\ell }\equiv
\sum_{\ell =1}^{\infty }J^{c;x^{\ell }},
\end{equation}%
where $c=x,y,z$. We refer to $J^{c;x^{\ell }}\equiv \sigma ^{c;x^{\ell
}}E_{x}^{\ell }$ as the $\ell $th-order current. The first term is the
linear response, the second term is the second-order nonlinear response, and
so on. If there is inversion symmetry in the system, the even-order
conductivities with even $\ell $ are prohibited. because%
\begin{equation}
J^{c}\mapsto -J^{c},\qquad E_{x}\mapsto -E_{x}
\end{equation}%
under inversion symmetry.

We assume that the applied electric field is alternating\ and monochromatic,%
\begin{equation}
E_{x}=E_{x}\left( \omega \right) e^{-i\omega t},  \label{monochro}
\end{equation}%
and investigate the $\ell $th-order bulk photovoltaic current.

\subsection{Second-order photovoltaic currents}

The second-order response has a form%
\begin{equation}
J^{c;x^{2}}\left( \omega _{1}+\omega _{2}\right) =\sigma ^{c;x^{2}}\left(
\omega _{1}+\omega _{2};\omega _{1},\omega _{2}\right) E_{x}\left( \omega
_{1}\right) E_{x}\left( \omega _{2}\right) .
\end{equation}%
We investigate the direct current generation, 
\begin{equation}
J^{c;x^{2}}\left( 0\right) =\sigma ^{c;x^{2}}\left( 0;\omega ,-\omega
\right) E_{x}\left( \omega \right) E_{x}\left( -\omega \right) .
\end{equation}%
In the following, we use the abbreviation $J^{c;x^{2}}\equiv
J^{c;x^{2}}\left( 0\right) $ and $\sigma ^{c;x^{2}}\left( \omega \right)
\equiv \sigma ^{c;x^{2}}\left( 0;\omega ,-\omega \right) $.

The conductivity of the injection current is in general given by the formula%
\cite{Sipe,JuanNC,Juan,Ave,AhnX,AhnNP,WatanabeInject,Okumura,Dai,EzawaVolta}%
\begin{equation}
\sigma _{\text{inject}}^{c;x^{2}}=-\tau \frac{2\pi e^{3}}{\hbar ^{2}}%
\sum_{n,m,\mathbf{k}}f_{nm}\Delta _{mn}^{c}r_{nm}^{x}r_{mn}^{x}\delta \left(
\omega _{mn}-\omega \right) ,  \label{Inject}
\end{equation}%
where $\tau $ is the relaxation time, $f_{nm}=f_{n}-f_{m}$ with $%
f_{n}=1/\left( \exp \left( \varepsilon _{n}-\mu \right) /k_{\text{B}%
}T+1\right) $ the Fermi distribution function for the band $n$, $\mu $ is
the chemical potential, $\varepsilon _{n}$ is the energy of the band $n$, 
\begin{equation}
r_{mn}^{x}=\left\langle m\right\vert i\partial _{k_{x}}\left\vert
n\right\rangle
\end{equation}%
is the Berry connection, 
\begin{equation}
\Delta _{mn}^{c}=v_{m}^{c}-v_{n}^{c}
\end{equation}%
is the the difference of the velocities defined by 
\begin{equation}
v_{m}^{c}=\frac{1}{\hbar }\left\langle m\right\vert \partial
_{k_{c}}H\left\vert m\right\rangle
\end{equation}%
with the Hamiltonian $H$, and $\omega _{nm}\equiv \left( \varepsilon
_{n}-\varepsilon _{m}\right) /\hbar $. The injection current is induced when
the velocities are different ($\Delta _{mn}^{c}\neq 0$) between the
conduction band $n$ and the valence band\textsl{\ }$m$ along the $c$
direction.

The shift current is in general given by the formula\cite%
{Young,Young2,Ave,Kraut,Baltz,Sipe,Juan,AhnX,MorimotoScAd,Kim,Barik,AhnNP,WatanabeInject,Dai,Yoshida,EzawaVolta}%
\begin{equation}
\sigma _{\text{shift}}^{c;x^{2}}=-\frac{\pi e^{3}}{\hbar ^{2}}\sum_{n,m,%
\mathbf{k}}f_{nm}R_{mn}^{c,x}r_{nm}^{x}r_{mn}^{x}\delta \left( \omega
_{mn}-\omega \right) ,  \label{Shift}
\end{equation}%
where 
\begin{equation}
R_{mn}^{c,x}=r_{mm}^{x}-r_{nn}^{x}+i\partial _{k_{c}}\log r_{mn}^{x}
\label{shift}
\end{equation}%
is the shift vector\cite{Sipe}. The shift vector is gauge invariant although
the Berry connection is not gauge invariant. The shift vector describes the
difference of the mean position of the Wannier function between two bands $m$
and $n$. The integrand is rewritten as%
\begin{equation}
R_{mn}^{c,x}r_{nm}^{x}r_{mn}^{x}=ir_{mn}^{x}r_{nm,c}^{x},
\end{equation}%
where we have defined the covariant derivative%
\begin{equation}
\nabla _{k_{c}}r_{nm}^{x}\equiv r_{nm,c}^{x}\equiv \frac{\partial r_{nm}^{x}%
}{\partial k_{c}}-ir_{nm}^{x}\left( r_{nn}^{x}-r_{mm}^{x}\right) .
\end{equation}%
The shift current is induced when the mean positions are different ($%
R_{mn}^{c,x}\neq 0$) between the conduction band $n$ and the valence band $m$%
.

In the following, we concentrate on the longitudinal conductivities by
setting $c=x$.

\subsection{Jerk current and snap current}

The Jerk current is a third-order bulk photocurrent\cite%
{Jerk,JerkComment,JerkReply}. It is calculated from%
\begin{equation}
\frac{d^{2}J_{\text{jerk}}^{x;x^{3}}}{dt^{2}}=\frac{2e}{V}\sum_{n}\frac{%
df_{n}}{dt}\frac{dv_{n}^{x}}{dt}E_{x}\left( \omega \right) E_{x}\left(
-\omega \right) E_{x}\left( 0\right) .
\end{equation}%
By assuming that the right-hand terms are constant, this equation is solved
as%
\begin{equation}
J_{\text{jerk}}^{x;x^{3}}\equiv \sigma _{\text{jerk}}^{x;x^{3}}E_{x}\left(
\omega \right) E_{x}\left( -\omega \right) E_{x}\left( 0\right) ,
\end{equation}%
where%
\begin{equation}
\frac{\sigma _{\text{jerk}}^{x;x^{3}}}{\sigma _{\text{jerk}}^{0}}=\tau
^{2}\sum_{n,m,\mathbf{k}}f_{nm}\frac{\partial ^{2}\omega _{mn}}{\partial
k_{x}^{2}}r_{mn}^{x}r_{nm}^{x}\delta \left( \omega _{mn}-\omega \right) ,
\end{equation}%
with $\sigma _{\text{jerk}}^{0}\equiv 2\pi e^{4}/\hbar ^{3}V$. It is
proportional to $\tau ^{2}$\cite{Jerk,AhnNP}.

The snap current is a fourth-order bulk photocurrent\cite{snap}. It is
calculated from%
\begin{equation}
\frac{d^{3}J_{\text{snap}}^{x;x^{4}}}{dt^{3}}=\frac{3e}{V}\sum_{n}\frac{%
df_{n}}{dt}\frac{d^{2}v_{n}^{x}}{dt^{2}}E_{x}\left( \omega \right)
E_{x}\left( -\omega \right) \left[ E_{x}\left( 0\right) \right] ^{2}.
\end{equation}%
By assuming that the right-hand terms are constant, this equation is solved
as%
\begin{equation}
J_{\text{snap}}^{x;x^{4}}\equiv \sigma _{\text{snap}}^{x;x^{4}}E_{x}\left(
\omega \right) E_{x}\left( -\omega \right) \left[ E_{x}\left( 0\right) %
\right] ^{2}
\end{equation}%
where%
\begin{equation}
\frac{\sigma _{\text{snap}}^{x;x^{4}}}{\sigma _{\text{snap}}^{0}}=\tau
^{3}\sum_{n,m,\mathbf{k}}f_{nm}\frac{\partial ^{3}\omega _{mn}}{\partial
k_{x}^{3}}r_{mn}^{x}r_{nm}^{x}\delta \left( \omega _{mn}-\omega \right) ,
\end{equation}%
with $\sigma _{\text{snap}}^{0}\equiv 3\pi e^{5}/\hbar ^{4}V$. It is
proportional to $\tau ^{3}$.

\subsection{Higher-order photovoltaic currents}

We generalize the injection and shift currents to the $\ell $-th order
photovoltaic currents. We analyze the $\ell $-th order photocurrent in the
form of%
\begin{equation}
J^{x;x^{\ell }}=\sigma ^{x;x^{\ell }}E_{x}\left( \omega \right) E_{x}\left(
-\omega \right) \left[ E_{x}\left( 0\right) \right] ^{\ell -2},
\end{equation}%
with $\ell -2$ components being static fields $E_{x}\left( 0\right) $. We
focus on the two-band model with $n=\pm $\ in the following.

\subsubsection{Higher-order injection currents}

We generalize the injection current to the $\ell $-th order photovoltaic
currents. The equation of motion of electrons is%
\begin{equation}
\hbar \frac{d\mathbf{k}}{dt}=-e\frac{\partial \mathbf{A}_{0}}{\partial t},
\end{equation}%
where $\mathbf{A}_{0}=\left( A_{0},0,0\right) $ is the vector potential. The
static electric field is obtained from the vector potential as%
\begin{equation}
E_{x}\left( 0\right) \equiv -\frac{\partial A_{0}}{\partial t},
\end{equation}%
or $A_{0}=-E_{x}\left( 0\right) t+$constant. The Bloch velocity under
electric field is given by the minimal substitution,%
\begin{equation}
v_{n}\left( k_{x}\right) \rightarrow v_{n}\left( k_{x}-eA_{0}/\hbar \right) .
\end{equation}%
We expand it in powers of $E_{x}\left( 0\right) $, and obtain%
\begin{equation}
\frac{d^{\ell }v_{n}^{x}}{dt^{\ell }}=\left( \frac{e}{\hbar }\right) ^{\ell }%
\frac{\partial ^{\ell +1}\omega _{n}}{\partial k_{x}^{\ell +1}}\left[
E_{x}\left( 0\right) \right] ^{\ell },
\end{equation}%
where we have used%
\begin{equation}
\frac{dv_{n}^{x}}{dt}=\frac{dk_{x}}{dt}\frac{dv_{n}^{x}}{dk_{x}}=-\left( 
\frac{e}{\hbar }\right) \frac{\partial A_{0}}{\partial t}\frac{d^{2}\omega
_{n}}{dk_{x}^{2}}=\frac{e}{\hbar }\frac{d^{2}\omega _{n}}{dk_{x}^{2}}%
E_{x}\left( 0\right) .
\end{equation}%
The current is given by%
\begin{equation}
J=\frac{e}{V}\sum_{n,\mathbf{k}}f_{n}v_{n}^{x},  \label{Jfv}
\end{equation}%
with the velocity along the $x$ direction%
\begin{equation}
v_{n}^{x}\equiv \frac{\partial \omega _{n}}{\partial k_{x}}.
\end{equation}%
The injection current originates in $dJ^{x;x^{2}}/dt$, the jerk current
originates in $d^{2}J^{x;x^{3}}/dt^{2}$\ and the snap current originates in $%
d^{3}J^{x;x^{4}}/dt^{3}$.

The $\ell $-th order injection current is derived from Eq.(\ref{Jfv}) as%
\begin{equation}
\frac{d^{\ell -1}J_{\text{inject}}^{x;x^{\ell }}}{dt^{\ell -1}}=\frac{e}{V}%
\sum_{n}\sum_{s=0}^{\ell -1}\left( 
\begin{array}{c}
\ell -1 \\ 
s%
\end{array}%
\right) \frac{d^{s}f_{n}}{dt^{s}}\frac{d^{\ell -1-s}v_{n}^{x}}{dt^{\ell -1-s}%
}.  \label{JEll}
\end{equation}%
The term proportional to $E_{x}\left( \omega \right) E_{x}\left( -\omega
\right) \left[ E_{x}\left( 0\right) \right] ^{\ell -2}$ is given by taking
the terms with $s=1$ from Eq.(\ref{JEll}) as%
\begin{equation}
\frac{d^{\ell -1}J_{\text{inject}}^{x;x^{\ell }}}{dt^{\ell -1}}=\left( \ell
-1\right) \frac{e}{V}\sum_{n}\frac{df_{n}}{dt}\frac{d^{\ell -2}v_{n}^{x}}{%
dt^{\ell -2}}.  \label{JElll}
\end{equation}%
The Fermi golden rule reads%
\begin{equation}
\frac{df_{\pm }}{dt}=\pm \frac{2\pi e^{2}}{\hbar ^{2}}\left\vert \mathbf{E}%
\left( \omega \right) \cdot \mathbf{r}_{+-}\right\vert \delta \left( \omega
_{+-}-\omega \right) .
\end{equation}%
Inserting it into Eq.(\ref{JElll}), we have%
\begin{align}
& \frac{d^{\ell -1}J_{\text{inject}}^{x;x^{\ell }}}{dt^{\ell -1}}  \notag \\
& =\frac{\left( \ell -1\right) 2\pi e^{2}}{\hbar ^{2}}\frac{e}{V}\left( 
\frac{e}{\hbar }\right) ^{\ell -2}r_{-+}^{x}r_{+-}^{x}\delta \left( \omega
_{+-}-\omega \right)  \notag \\
& \times \frac{\partial ^{\ell +1}\omega _{+-}}{\partial k_{x}^{\ell +1}}%
E_{x}\left( \omega \right) E_{x}\left( -\omega \right) \left[ E_{x}\left(
0\right) \right] ^{\ell -2}.
\end{align}%
By assuming a monochromic oscillation $J^{x;x^{\ell }}\propto e^{-i\omega
_{0}t}$, we obtain the $\ell $th-order current $J^{x;x^{\ell }}$,%
\begin{align}
J_{\text{inject}}^{x;x^{\ell }}=& \frac{\ell -1}{\left( i\omega _{0}+1/\tau
\right) ^{\ell -1}}\frac{2\pi e^{\ell +1}}{\hbar ^{\ell }V}%
r_{-+}^{x}r_{+-}^{x}\delta \left( \omega _{+-}-\omega \right)  \notag \\
& \times \frac{\partial ^{\ell -1}\omega _{+-}}{\partial k_{x}^{\ell -1}}%
E_{x}\left( \omega \right) E_{x}\left( -\omega \right) \left[ E_{x}\left(
0\right) \right] ^{\ell -2},
\end{align}%
where we have introduced a cut off by the relaxation time $\tau $. When we
concentrate on the direct current component $\omega _{0}=0$, it is
proportional to $\tau ^{\ell -1}$. The longitudinal component is 
\begin{equation}
\frac{\sigma _{\text{inject}}^{x;x^{\ell }}}{\sigma _{\text{inject}}^{0}}%
=\left( \ell -1\right) \tau ^{\ell -1}\sum_{\mathbf{k}}f_{-+}\frac{\partial
^{\ell -1}\omega _{+-}}{\partial k_{x}^{\ell -1}}r_{-+}^{x}r_{+-}^{x}\delta
\left( \omega _{+-}-\omega \right) ,
\end{equation}%
with $\sigma _{\text{inject}}^{\left( \ell \right) }\equiv 2\pi \left( \ell
-1\right) e^{\ell +1}/\hbar ^{\ell }V$. This formula is reduced to the
injection current ($\ell =2$), the jerk current ($\ell =3$) and the snap
current ($\ell =4$). They are named the crackle current ($\ell =5$), the pop
current ($\ell =6$), the lock current ($\ell =7$), the drop current ($\ell
=8 $), the shot current ($\ell =9$) and the put current ($\ell =10$)\cite%
{Matt,snap}.

\subsubsection{Higher-order shift currents}

The shift current originates in a quantum interference\cite{snap} between
the oscillation of $\rho _{mn}$\ and the oscillation of the dipole velocity $%
E_{x}r_{nm;x}^{x}$. Hence, it is natural to define a higher-order shift
current by%
\begin{equation}
J_{\text{shift}}^{x;x^{\ell }}=-\frac{e^{2}}{\hbar V}\sum_{\mathbf{k}%
}E_{x}r_{+-;x}^{x}\rho _{+-}^{\left( \ell -1\right) },
\end{equation}%
where $\rho _{mn}^{\left( \ell -1\right) }$ is the $\ell $-th order solution
of the density matrix with the von-Neumann equation\cite{Sipe,Ave,snap},%
\begin{eqnarray}
&&\frac{\partial \rho _{mn}}{\partial t}+i\omega _{mn}\rho _{mn}  \notag \\
&=&\frac{e}{i\hbar }\sum_{s}E_{x}\left( \rho _{ml}r_{sn}^{x}-r_{ms}^{x}\rho
_{sn}\right) -\frac{e}{\hbar }E_{x}\rho _{mn;x}.
\end{eqnarray}%
For the two-band system, it is rewritten as%
\begin{equation}
i\left( \omega _{+-}-\omega _{0}\right) \rho _{+-}=\frac{e}{i\hbar }%
E_{x}r_{+-}^{x}\left( \rho _{++}-\rho _{--}\right) -\frac{e}{\hbar }%
E_{x}\rho _{+-;x},
\end{equation}%
where we have used $r_{nn}^{x}=0$. This definition recovers properly the
definition of the shift current for $\ell =2$.

The zeroth-order solution is given by%
\begin{equation}
\rho _{\pm }^{\left( 0\right) }=f_{\pm },
\end{equation}%
because electric field is not applied. The first-order solution is given by%
\begin{align}
\rho _{+-}^{\left( 1\right) } &=\frac{ie}{\hbar \left( \omega _{+-}-\omega
_{0}\right) }E_{x}r_{+-}^{x}\left( \rho _{++}^{\left( 0\right) }-\rho
_{--}^{\left( 0\right) }\right) e^{-i\omega _{0}t}  \notag \\
&=\frac{ie}{\hbar \left( \omega _{+-}-\omega _{0}\right) }%
E_{x}r_{+-}^{x}f_{-+}e^{-i\omega _{0}t}.
\end{align}%
Especially, we have $\rho _{++}^{\left( 1\right) }=0$ because $f_{++}\equiv
f_{+}-f_{+}=0$.

The recursive equation reads%
\begin{align}
\rho _{+-}^{\left( \ell \right) }& =\frac{e}{i\hbar }E_{x}r_{+-}^{x}\left(
\rho _{++}^{\left( \ell -1\right) }-\rho _{--}^{\left( \ell -1\right)
}\right) +\frac{ieE_{x}}{\hbar \left( \omega _{+-}-\omega _{0}\right) }\rho
_{+-;x}^{\left( \ell -1\right) }  \notag \\
& =\frac{ieE_{x}}{\hbar \left( \omega _{+-}-\omega _{0}\right) }\nabla
_{k_{x}}\rho _{+-}^{\left( \ell -1\right) }  \notag \\
& =\left( \frac{ieE_{x}}{\hbar \left( \omega _{+-}-\omega _{0}\right) }%
\right) ^{\ell -1}\nabla _{k_{x}}^{\ell -1}\rho _{+-}^{\left( 1\right) } 
\notag \\
& =\left( \frac{ieE_{x}}{\hbar \left( \omega _{+-}-\omega _{0}\right) }%
\right) ^{\ell -1}E_{x}f_{-+}\nabla _{k_{x}}^{\ell -1}r_{+-}^{x},
\end{align}%
where we have used $\rho _{++}^{\left( \ell -1\right) }=\rho _{--}^{\left(
\ell -1\right) }$. Hence, the $\ell $-th order shift current is obtained as%
\begin{align}
J_{\text{shift}}^{x;x^{\ell }}& =-\frac{e^{2}}{\hbar V}\sum_{\mathbf{k}%
}E_{x}r_{+-;x}^{x}\rho _{+-}^{\left( \ell -1\right) }  \notag \\
& =-\frac{e^{2}}{\hbar V}\left( \frac{ieE_{x}}{\hbar \left( \omega
_{+-}-\omega _{0}\right) }\right) ^{\ell -2}e^{-i\omega _{0}t}  \notag \\
& \times \sum_{\mathbf{k}}r_{+-;x}^{x}f_{-+}\nabla _{k_{x}}^{\ell
-2}r_{+-}^{x}E_{x}^{\ell }.
\end{align}%
We use the monochromatic condition (\ref{monochro}) and study the direct
current component%
\begin{align}
J_{\text{shift}}^{x;x^{\ell }}=& \sigma _{\text{shift}}^{x;x^{\ell }}\left(
0;\omega ,-\omega ,0,0,\cdots ,0\right)  \notag \\
& \times E_{x}\left( \omega \right) E_{x}\left( -\omega \right) \left[
E_{x}\left( 0\right) \right] ^{\ell -2}.
\end{align}%
Then, the $\ell $-th order shift current is obtained as 
\begin{align}
J_{\text{shift}}^{x;x^{\ell }}=& -\frac{e^{2}}{\hbar V}\left( \frac{ie}{%
\hbar \omega }\right) ^{\ell -2}\sum_{\mathbf{k}}r_{+-;x}^{x}\nabla
_{k_{x}}^{\ell -2}r_{+-}^{x}  \notag \\
& \times f_{-+}E_{x}\left( \omega \right) E_{x}\left( -\omega \right) \left[
E_{x}\left( 0\right) \right] ^{\ell -2}\delta \left( \omega _{+-}-\omega
\right) .
\end{align}%
It is independent of $\tau $. The $\ell $-th order shift conductivity is
given by 
\begin{equation}
\frac{\sigma _{\text{shift}}^{x;x^{\ell }}}{\sigma _{\text{inject}}^{\left(
\ell \right) }}=\sum_{\mathbf{k}}f_{-+}r_{+-;x}^{x}\nabla _{k_{x}}^{\ell
-2}r_{+-}^{x}\delta \left( \omega _{+-}-\omega \right) ,
\end{equation}%
with 
\begin{equation}
\sigma _{\text{inject}}^{\left( \ell \right) }\equiv -\frac{e^{3}}{\hbar
^{2}V}\left( \frac{ie}{\hbar \omega }\right) ^{\ell -2}.
\end{equation}%
It recovers an ordinary shift current for $\ell =2$.

\section{Quantum geometry}

Quantum geometry is differential geometry based on the normalized complex
vectors in quantum mechanics. The quantum geometric tensor is defined by\cite%
{Provost,Ma}%
\begin{equation}
\mathcal{F}_{mn}^{\mu \nu }\left( \mathbf{k}\right) =\left\langle \partial
_{k_{\mu }}\psi _{m}\left( \mathbf{k}\right) \right\vert 1-\mathcal{P}\left( 
\mathbf{k}\right) \left\vert \partial _{k_{\nu }}\psi _{n}\left( \mathbf{k}%
\right) \right\rangle ,
\end{equation}%
where $\mu =x,y,z$, and 
\begin{equation}
\mathcal{P}\left( \mathbf{k}\right) \equiv \sum_{n\in \text{Occupied}%
}\left\vert u_{n}\left( \mathbf{k}\right) \right\rangle \left\langle
u_{n}\left( \mathbf{k}\right) \right\vert
\end{equation}%
is the projection operator to the occupied band.

In the two-band model, its diagonal component is simply given by the quantum
metric%
\begin{align}
\mathcal{F}_{--}^{\mu \nu }\left( \mathbf{k}\right) & =\left\langle \partial
_{k_{\mu }}\psi _{-}(\mathbf{k})\left\vert \psi _{+}(\mathbf{k}%
)\right\rangle \right. \left. \left\langle \psi _{+}(\mathbf{k})\right\vert
\partial _{k_{\nu }}\psi _{-}(\mathbf{k})\right\rangle  \notag \\
& =r_{-+}^{\mu \ast }\left( \mathbf{k}\right) r_{+-}^{\nu }\left( \mathbf{k}%
\right) =g_{\mu \nu }\left( \mathbf{k}\right) .
\end{align}%
In general, the Hamiltonian of the two-band model is given by 
\begin{equation}
H\left( \mathbf{k}\right) =\sigma _{0}h_{0}\left( \mathbf{k}\right) +\mathbf{%
\sigma }\cdot \mathbf{h}\left( \mathbf{k}\right) ,  \label{Hamil2}
\end{equation}%
where $\mathbf{h}\left( \mathbf{k}\right) $ is parametrized as%
\begin{equation}
\mathbf{n}\left( \mathbf{k}\right) \equiv \mathbf{h}/|\mathbf{h}|=\left(
\sin \theta \cos \phi ,\sin \theta \sin \phi ,\cos \theta \right) ,
\label{dVector}
\end{equation}%
and $\sigma _{j}$ is the Pauli matrix with $j=x,y,z$. The eigenfunction of
the Hamiltonian (\ref{Hamil2}) is given by%
\begin{equation}
\psi _{\pm }\left( \mathbf{k}\right) =\left( 
\begin{array}{c}
\pm e^{-i\phi }\sin \frac{\theta }{2} \\ 
\cos \frac{\theta }{2}%
\end{array}%
\right) .
\end{equation}%
Then, the Berry connection is calculated as%
\begin{align}
r_{\mu }^{-+}\left( \mathbf{k}\right) =& -i\left\langle \psi _{-}\left( 
\mathbf{k}\right) \right\vert \partial _{\mu }\psi _{+}\left( \mathbf{k}%
\right) \rangle  \notag \\
=& -i\left( -e^{-i\phi }\sin \frac{\theta }{2},\cos \frac{\theta }{2}\right)
\notag \\
& \times \left( 
\begin{array}{c}
-ie^{-i\phi }\partial _{\mu }\phi \sin \frac{\theta }{2}+\frac{\partial
_{\mu }\theta }{2}e^{-i\phi }\cos \frac{\theta }{2} \\ 
-\frac{\partial _{\mu }\theta }{2}\sin \frac{\theta }{2}%
\end{array}%
\right)  \notag \\
=& \partial _{\mu }\phi \sin ^{2}\frac{\theta }{2}-i\frac{\partial _{\mu
}\theta }{2}\cos \frac{\theta }{2}\sin \frac{\theta }{2}+i\frac{\partial
_{\mu }\theta }{2}\cos \frac{\theta }{2}\sin \frac{\theta }{2}  \notag \\
=& \partial _{\mu }\phi \sin ^{2}\frac{\theta }{2}.
\end{align}%
It is real, $\text{Im}r_{-+}^{\mu }\left( \mathbf{k}\right) =0$, because $%
\theta $ and $\phi $ are real. Then, we have%
\begin{equation}
r_{+-}^{\mu }r_{-+}^{\nu }=g_{\mu \nu }.
\end{equation}%
We use this relation to rewrite the injection current, the shift current and
the jerk current in terms of the quantum metric.

The injection current is rewritten in terms of the quantum metric\cite%
{AhnX,AhnNP},%
\begin{equation}
\sigma _{\text{inject}}^{x;x^{2}}=-\tau \frac{2\pi e^{3}}{\hbar ^{2}}\sum_{%
\mathbf{k}}f_{-+}\Delta _{+-}^{x}g_{xx}\delta \left( \omega _{+-}-\omega
\right) ,
\end{equation}

The shift current is rewritten in terms of the quantum metric\cite%
{WatanabeInject}%
\begin{equation}
\sigma _{\text{shift}}^{x;x^{2}}=-\frac{\pi e^{3}}{\hbar ^{2}}\sum_{\mathbf{k%
}}f_{-+}R_{+-}^{x,x}g_{xx}\delta \left( \omega _{+-}-\omega \right) .
\label{Rgxx}
\end{equation}%
In a similar way, the higher-order injection current is rewritten in term of
the quantum metric as%
\begin{equation}
\frac{\sigma _{\text{inject}}^{x;x^{\ell }}}{\sigma _{\ell }^{0}}=\sum_{%
\mathbf{k}}f_{-+}\frac{\partial ^{\ell -1}\omega _{+-}}{\partial k_{x}^{\ell
-1}}g_{xx}\delta \left( \omega _{+-}-\omega \right) .
\end{equation}

There is another representation\cite{AhnNP} equivalent to Eq.(\ref{Rgxx}), 
\begin{equation}
\sigma _{\text{shift}}^{x;x^{2}}=-\frac{\pi e^{3}}{\hbar ^{2}}\sum_{\mathbf{k%
}}f_{-+}iC_{xxx}^{-+}\delta \left( \omega _{+-}-\omega \right) ,
\end{equation}%
where the quantum connection is defined by%
\begin{align}
C_{xxx}^{mn} &\equiv \left( \hat{e}_{x}^{mn},\nabla _{x}\hat{e}%
_{x}^{mn}\right) \equiv r_{nm}^{x}r_{mn,x}^{x}  \notag \\
&=r_{nm}^{x}\left[ \frac{\partial r_{nm}^{x}}{\partial k_{x}}%
-ir_{nm}^{x}\left( r_{nn}^{x}-r_{mm}^{x}\right) \right]  \label{QCone}
\end{align}%
with\cite{AhnNP}%
\begin{equation}
\hat{e}_{x}^{mn}\left( \mathbf{k}\right) \equiv r_{nm}^{x}\left\vert
u_{m}\left( \mathbf{k}\right) \right\rangle \left\langle u_{n}\left( \mathbf{%
k}\right) \right\vert .
\end{equation}

We generalize this relation by introducing the quantum connection to the
higher-order quantum connection defined by%
\begin{equation}
C_{xxxx}^{mn\left( \ell \right) }\equiv \left( \nabla _{x}^{\ell }\hat{e}%
_{x}^{mn},\nabla _{x}\hat{e}_{x}^{mn}\right) \equiv \frac{\nabla _{x}^{\ell
}r_{mn}^{x}}{\nabla k_{x}^{\ell }}r_{nm;x}^{x}.  \label{HCone}
\end{equation}%
By using it, the $\ell $-th order shift conductivity is expressed in terms
of it as%
\begin{equation}
\sigma _{\text{shift}}^{x;x^{\ell }}=-\frac{e^{3}}{\hbar ^{2}V}\left( \frac{%
ie}{\hbar \omega }\right) ^{\ell -2}\sum_{\mathbf{k}}f_{-+}C_{xxxx}^{+-%
\left( \ell \right) }\delta \left( \omega _{+-}-\omega \right) .
\end{equation}

\section{$p$-wave magnet}

In the following, we apply these results on photovoltaic currents to the
p-wave magnet as a typical example, where various formulas are analytically
obtained without using approximations. A prominent feature of the p-wave
magnet is that there is no second-order photovoltaic currents but there are
nonzero higher-order photovoltaic currents for $\ell \geq 3$.

We analyze the Hamiltonian given by%
\begin{equation}
H=H_{0}+H_{p}+H_{B}.  \label{TotalHamil}
\end{equation}%
The first term represents the kinetic energy of electrons,%
\begin{equation}
H_{0}=\frac{\hbar ^{2}k_{x}^{2}}{2m}\sigma _{0},
\end{equation}%
where $m$ is the electron mass, and $\sigma _{0}$ is $2\times 2$ identity
matrix. The second term%
\begin{equation}
H_{p}\left( \mathbf{k}\right) =J\left( \mathbf{s}\cdot \mathbf{\sigma }%
\right) k_{x}
\end{equation}%
describes the effect of the $p$-wave magnet on the electrons\cite%
{pwave,Maeda,EzawaPwave,Brek,GI,Edel,Elliptic}, where $J\mathbf{s\ }$is the N%
\'{e}el vector with $J$\ being its magnitude and $\mathbf{s}=\left( \sin
\Theta \cos \Phi ,\sin \Theta \sin \Phi ,\cos \Theta \right) $ being its
direction. We set $J_{x}=J\sin \Theta \cos \Phi $, $J_{y}=J\sin \Theta \sin
\Phi $ and $J_{z}=J\cos \Theta $. The third term is the magnetic field term,%
\begin{equation}
H_{B}=B\sigma _{z}.
\end{equation}%
Then, the Hamiltonian (\ref{TotalHamil}) is of the form of Eq.(\ref{Hamil2})
with%
\begin{align}
& h_{0}=\frac{\hbar ^{2}k_{x}^{2}}{2m}, & h_{x}& =Jk_{x}\sin \Theta \cos
\Phi ,  \notag \\
& h_{y}=Jk_{x}\sin \Theta \sin \Phi , & h_{z}& =Jk_{x}\cos \Theta +B.
\end{align}

\begin{figure}[t]
\centerline{\includegraphics[width=0.48\textwidth]{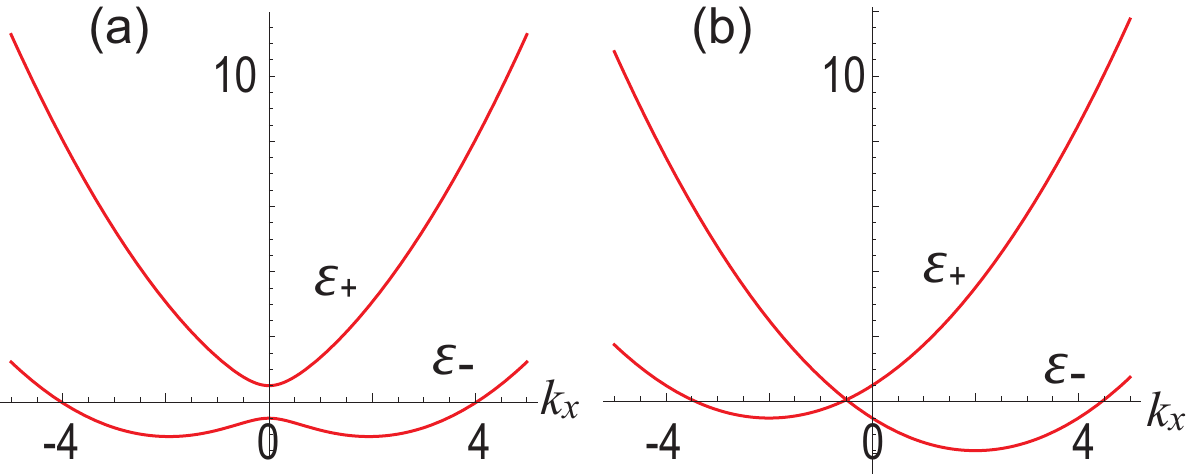}}
\caption{Energy spectrum in units of $\protect\varepsilon _{0}$. (a) The N%
\'{e}el vector is along the $x$ direction ($\Theta =\protect\pi /2;\Phi =0$%
). (b) The N\'{e}el vector is along the $z$ direction ($\Theta =0$). The
horizontal axis is $k_x$. We have set $J=\protect\varepsilon _{0}/k_{0}$, $%
\hbar ^{2}k_{0}^{2}/\left( 2m\right) =\protect\varepsilon _{0}/2$ and $B=0.5%
\protect\varepsilon _{0}$.}
\label{FigBand}
\end{figure}

There is inversion symmetry for $\Theta =\pi /2$,%
\begin{equation}
\sigma _{z}H_{p}\left( k_{x}\right) \sigma _{z}=H_{p}\left( -k_{x}\right) ,
\end{equation}%
where the N\'{e}el vector points to an in-plane direction.

The energy spectrum of the Hamiltonian (\ref{TotalHamil}) is given by%
\begin{equation}
\varepsilon _{\pm }=\frac{\hbar ^{2}k_{x}^{2}}{2m}\pm \sqrt{%
J^{2}k_{x}^{2}\sin ^{2}\Theta +\left( B+Jk_{x}\cos \Theta \right) ^{2}},
\end{equation}%
which demonstrates the p-wave spin splitting. By solving $\omega _{+-}\left(
k_{\pm }\right) -\omega =0$, we obtain%
\begin{equation}
k_{\pm }=-\frac{-BJ_{z}}{J}\pm \frac{\sqrt{J^{2}\omega
^{2}-4B^{2}J_{\parallel }^{2}}}{2J^{2}}.
\end{equation}%
The band structure is shown in Fig.\ref{FigBand}(a) and in Fig.\ref{FigBand}%
(b) when the N\'{e}el vector is taken along the $x$\ axis and along the $z$\
axis, respectively.

The quantum metric for the two-band system is explicitly given by\cite%
{Matsuura,Gers,Onishi,WChen2024,EzawaQG,Oh}%
\begin{equation}
g_{\mu \nu }\left( \mathbf{k}\right) =\frac{1}{2}\left( \partial _{k_{\mu }}%
\mathbf{n}\right) \cdot \left( \partial _{k_{\nu }}\mathbf{n}\right) ,
\label{Gmn}
\end{equation}%
with the normalized vector (\ref{dVector}). Explicitly, the quantum metric
is obtained as 
\begin{equation}
g_{xx}=\frac{B^{2}J_{\parallel }^{2}}{2\left(
B^{2}+J^{2}k_{x}^{2}+2BJ_{z}\right) ^{2}}  \label{gxx}
\end{equation}%
with $J_{\parallel }^{2}\equiv J_{x}^{2}+J_{y}^{2}$.

The quantum connection (\ref{QCone}) is calculated as%
\begin{equation}
C_{xxx}^{-+}=\frac{B^{2}J_{\parallel }^{2}\left( J^{2}k_{x}+BJ_{z}\right) }{%
\left( B^{2}+J^{2}k_{x}^{2}+2BJ_{z}\right) ^{3}},  \label{HeCone0}
\end{equation}%
and the higher-order Hermitian connection (\ref{HCone}) is calculated as%
\begin{align}
C_{xxx}^{-+\left( 1\right) }& =\frac{B^{2}J_{\parallel }^{2}\left(
J^{2}k_{x}+BJ_{z}\right) ^{2}}{\left( B^{2}+J^{2}k_{x}^{2}+2BJ_{z}\right)
^{4}}, \\
C_{xxx}^{-+\left( 2\right) }& =\frac{B^{2}J_{\parallel }^{2}\left(
J^{2}k_{x}+BJ_{z}\right) }{\left( B^{2}+J^{2}k_{x}^{2}+2BJ_{z}\right) ^{5}} 
\notag \\
& \times \left( 3J^{4}k_{x}^{2}+3B^{2}J_{z}^{2}-B^{2}J_{\parallel
}^{2}+6BJ^{2}J_{z}k_{x}\right) ,
\end{align}%
and so on. We find it is proportional to%
\begin{equation}
C_{xxx}^{-+\left( \ell \right) }\propto \frac{B^{2}J_{\parallel }^{2}\left(
J^{2}k_{x}+BJ_{z}\right) }{\left( B^{2}+J^{2}k_{x}^{2}+2BJ_{z}\right) ^{\ell
+3}}.  \label{HeCone}
\end{equation}

At the zero temperature, the Fermi distribution difference is reduced to be%
\begin{equation}
f_{-+}=\theta \left( \frac{\omega }{2}-\frac{\hbar ^{2}k_{x}^{2}}{2m}+\mu
\right) ,
\end{equation}%
where $\theta \left( x\right) $ is the step function $\theta \left( x\right)
=1$ for $x>0$ and $\theta \left( x\right) =0$ for $x<0$. In the following,
we set $f_{-+}=1$ a the zero temperature. The validity of this assumption is
discussed in the \ Sec.IV.E.

\begin{figure}[t]
\centerline{\includegraphics[width=0.48\textwidth]{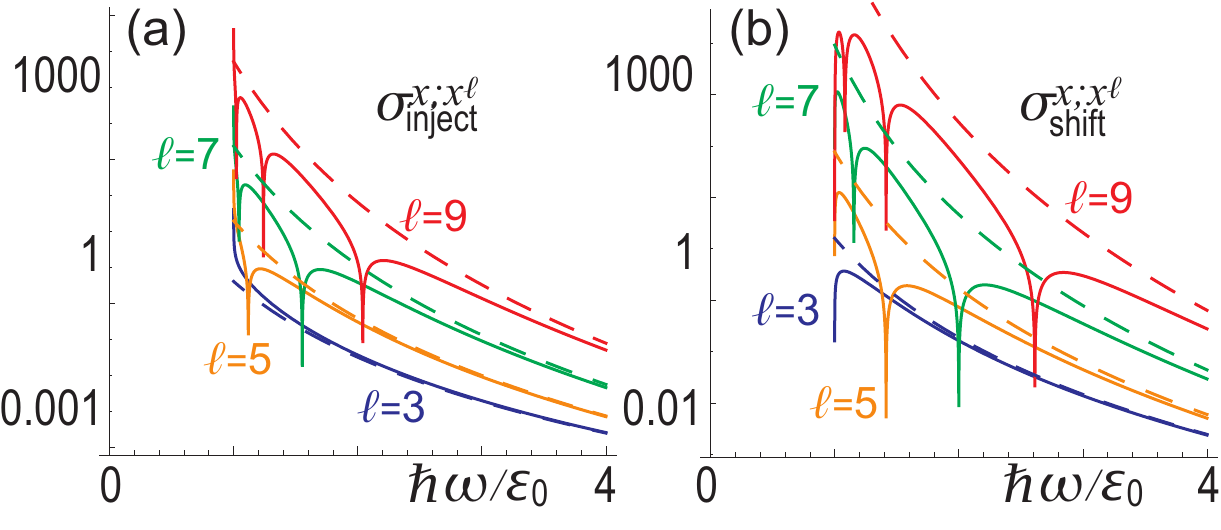}}
\caption{ (a) Higher-order injection currents log$_{10}\left\vert \protect%
\sigma _{\text{inject}}^{x;x^{\ell }}/\protect\sigma _{\text{inject}%
}^{\left( \ell \right) }\right\vert $ in units of $\protect\sigma _{\text{%
inject}}^{\left( \ell \right) }$. (b) Higher-order shift currents log$%
_{10}\left\vert \protect\sigma _{\text{shift}}^{x;x^{\ell }}/\protect\sigma %
_{\text{shift}}^{\left( \ell \right) }\right\vert $ in units of $\protect%
\sigma _{\text{shift}}^{\left( \ell \right) }$. Blue curves indicate $\ell
=3 $, orange curves indicates $\ell =5$, green curves indicate $\ell =7$ and
red curves indicate $\ell =9$. Dashed curves indicate asymptotic
behaviors for large $\protect\omega $. We have set $J=\protect\varepsilon %
_{0}/k_{0}$, $\hbar ^{2}k_{0}^{2}/\left( 2m\right) =\protect\varepsilon %
_{0}/2$ and $B=0.5\protect\varepsilon _{0}$. }
\label{FigHighPhoto}
\end{figure}

\subsection{Injection current}

The injection current at the zero temperature is calculated in the
Hamiltonian (\ref{TotalHamil}) as%
\begin{equation}
\sigma _{\text{inject}}^{x;x^{2}}=-\tau \frac{2\pi e^{3}}{\hbar ^{2}}%
\sum_{k_{x}}\Delta _{+-}^{x}g_{xx}\frac{\delta \left( k_{x}-k_{\pm }\right) 
}{2\left\vert \partial _{k_{x}}\omega _{+-}\right\vert }.
\end{equation}%
By using the Hamiltonian (\ref{TotalHamil}), we have%
\begin{align}
\sum_{k_{x}}& \Delta _{+-}^{x}g_{xx}\frac{\delta \left( k_{x}-k_{0}\right) }{%
2\left\vert \partial _{k_{x}}\omega _{+-}\right\vert }  \notag \\
=& \sum_{k_{x}=k_{\pm }}\frac{2\left( J^{2}k_{x}+BJ_{z}\right) }{\sqrt{%
B^{2}+J^{2}k_{x}^{2}+2BJ_{z}}}\frac{B^{2}J_{\parallel }^{2}}{4\left(
B^{2}+J^{2}k_{x}^{2}+2BJ_{z}\right) ^{2}}  \notag \\
& \times \frac{\sqrt{B^{2}+J^{2}k_{x}^{2}+2BJ_{z}}}{4\left\vert
J^{2}k_{x}+BJ_{z}\right\vert }  \notag \\
=& \sum_{k_{x}=k_{\pm }}\frac{\left( J^{2}k_{x}+BJ_{z}\right) }{2\left\vert
J^{2}k_{x}+BJ_{z}\right\vert }\frac{B^{2}J_{\parallel }^{2}}{4\left(
B^{2}+J^{2}k_{x}^{2}+2BJ_{z}\right) ^{2}},
\end{align}%
where we have used 
\begin{equation}
\Delta _{+=}^{x}=\frac{2\left( J^{2}k_{x}+BJ_{z}\right) }{\sqrt{%
B^{2}+J^{2}k_{x}^{2}+2BJ_{z}}}.
\end{equation}%
By using%
\begin{equation}
B^{2}+J^{2}k_{0}^{2}+2BJ_{z}=\frac{\omega ^{2}}{4},  \label{Bk0}
\end{equation}%
and%
\begin{equation}
2\left( J^{2}k_{0}+BJ_{z}\right) =\eta \sqrt{J^{2}\omega
^{2}-4B^{2}J_{\parallel }^{2}},  \label{Jk0}
\end{equation}%
we have%
\begin{align}
& \sum_{k_{x}}\Delta _{mn}^{x}g_{xx}\delta \left( \omega _{+-}-\omega \right)
\notag \\
& =\sum_{\eta =\pm }\frac{\eta \sqrt{J^{2}\omega ^{2}-4B^{2}J_{\parallel
}^{2}}}{2\left\vert \eta \sqrt{J^{2}\omega ^{2}-4B^{2}J_{\parallel }^{2}}%
\right\vert }\frac{B^{2}J_{\parallel }^{2}}{4\left( \frac{\omega ^{2}}{4}%
\right) ^{2}}  \notag \\
& =\sum_{\eta =\pm }\eta \frac{2B^{2}J_{\parallel }^{2}}{\omega ^{4}}=0.
\end{align}%
Then, the injection current is exactly zero.

\subsection{Shift current}

The shift current at the zero temperature is zero in the Hamiltonian (\ref%
{TotalHamil}) because%
\begin{align}
\sigma _{\text{shift}}^{x;x^{2}}=& -\frac{\pi e^{3}}{\hbar ^{2}}%
\sum_{k_{x}}iC_{xxx}^{-+}\frac{\delta \left( k_{x}-k_{\pm }\right) }{%
2\left\vert \partial _{k_{x}}\omega _{+-}\right\vert }  \notag \\
=& -\frac{\pi e^{3}}{\hbar ^{2}}\sum_{n,m,k_{x}=k_{\pm }}i\frac{%
B^{2}J_{\parallel }^{2}\left( J^{2}k_{x}+BJ_{z}\right) }{\left(
B^{2}+J^{2}k_{x}^{2}+2BJ_{z}\right) ^{3}}  \notag \\
& \times \frac{\sqrt{B^{2}+J^{2}k_{x}^{2}+2BJ_{z}}}{4\left\vert
J^{2}k_{x}+BJ_{z}\right\vert }  \notag \\
=& -\frac{\pi e^{3}}{\hbar ^{2}}\sum_{n,m,\eta =\pm }\eta \frac{%
2iB^{2}J_{\parallel }^{2}}{\omega ^{5}}=0
\end{align}%
where we have used Eq.(\ref{HeCone0}).

\begin{figure}[t]
\centerline{\includegraphics[width=0.48\textwidth]{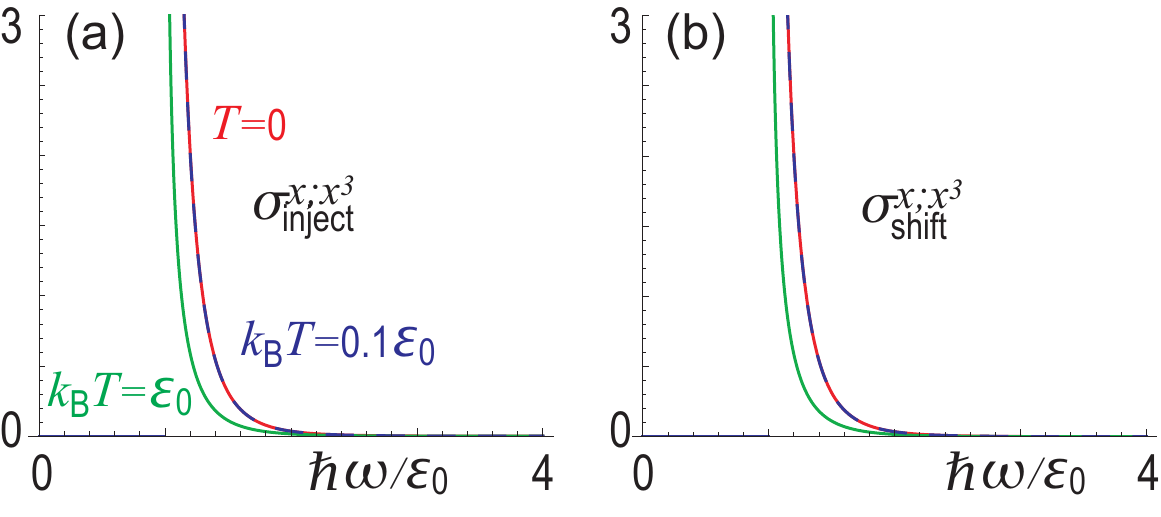}}
\caption{(a) Jerk current at finite temperature. (b) Third-order shift
current at finite temperature. Red dashed curves indicate the absolute zero
temperature $T=0$, blue dashed curves indicates $k_{\text{B}}T=0.1%
\protect\varepsilon _{0}$ and green curves indicate $k_{\text{B}}T=\protect%
\varepsilon _{0}$. We have set $\Theta =\protect\pi /2,$ $J=\protect%
\varepsilon _{0}/k_{0}$, $\hbar ^{2}k_{0}^{2}/\left( 2m\right) =\protect%
\varepsilon _{0}/2$ and $B=0.5\protect\varepsilon _{0}$. }
\label{FigJerkTemp}
\end{figure}
\begin{figure}[t]
\centerline{\includegraphics[width=0.48\textwidth]{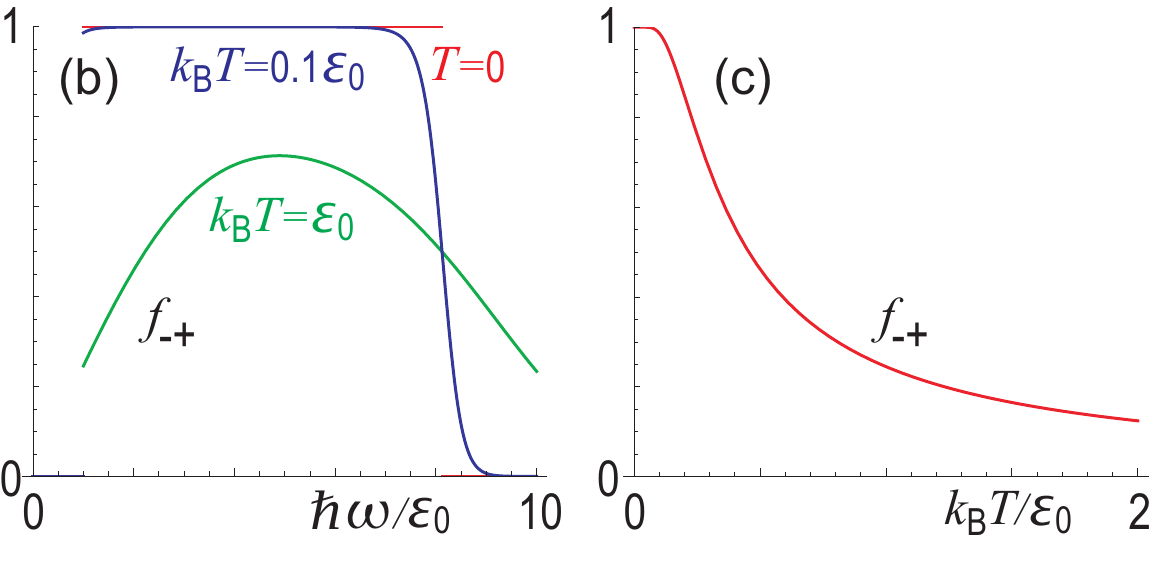}}
\caption{(a) $\protect\omega $ dependence of the Fermi distribution
difference $f_{-+}$. (b) Temperature dependence of the Fermi distribution
difference $f_{-+}$ at the optical band edge $\protect\omega =2B$. We have
set $J=\protect\varepsilon _{0}/k_{0}$, $\hbar ^{2}k_{0}^{2}/\left(
2m\right) =\protect\varepsilon _{0}/2$ and $B=0.5\protect\varepsilon _{0}$. }
\label{FigTemp}
\end{figure}

\subsection{Higher-order injection current}

\subsubsection{Jerk current}

Since the injection current and the shift current is zero, we study the jerk
current. The jerk current is calculated in the Hamiltonian (\ref{TotalHamil}%
) as%
\begin{equation}
\frac{\sigma _{\text{jerk}}^{x;x^{3}}}{\sigma _{\text{jerk}}^{\left(
3\right) }}=\sum_{k_{x}}f_{-+}\frac{\partial ^{2}\omega _{+-}}{\partial
k_{x}^{2}}g_{xx}\frac{\delta \left( k_{x}-k_{0}\right) }{2\left\vert
\partial _{k_{x}}\omega _{+-}\right\vert }.
\end{equation}%
By using the Hamiltonian (\ref{TotalHamil}), we have%
\begin{align}
\frac{\partial ^{2}\omega _{+-}}{\partial k_{x}^{2}}& \frac{g_{xx}}{%
2\left\vert \partial _{k_{x}}\omega _{+-}\right\vert }  \notag \\
=& \frac{2B^{2}J_{\parallel }^{2}}{\left(
B^{2}+J^{2}k_{x}^{2}+2BJ_{z}\right) ^{3/2}}\frac{B^{2}J_{\parallel }^{2}}{%
4\left( B^{2}+J^{2}k_{x}^{2}+2BJ_{z}\right) ^{2}}  \notag \\
& \times \frac{\sqrt{B^{2}+J^{2}k_{x}^{2}+2BJ_{z}}}{4\left\vert
J^{2}k_{x}+BJ_{z}\right\vert }  \notag \\
=& \frac{B^{4}J_{\parallel }^{4}}{8\left(
B^{2}+J^{2}k_{x}^{2}+2BJ_{z}\right) ^{3}\left\vert
J^{2}k_{x}+BJ_{z}\right\vert },
\end{align}%
where we have used%
\begin{equation}
\frac{\partial ^{2}\omega _{+-}}{\partial k_{x}^{2}}=\frac{%
2B^{2}J_{\parallel }^{2}}{\left( B^{2}+J^{2}k_{x}^{2}+2BJ_{z}\right) ^{3/2}},
\end{equation}%
and%
\begin{equation}
\partial _{k_{x}}\omega _{+-}=\frac{2\left( J^{2}k_{x}+BJ_{z}\right) }{\sqrt{%
B^{2}+J^{2}k_{x}^{2}+2BJ_{z}}},
\end{equation}%
for 
\begin{equation}
\omega >2\left\vert BJ_{\parallel }/J\right\vert \equiv \omega _{\text{cr}}.
\end{equation}%
By using (\ref{Bk0}) and (\ref{Jk0}), we have%
\begin{equation}
\left. \frac{\partial ^{2}\omega _{+-}}{\partial k_{x}^{2}}\frac{g_{xx}}{%
2\left\vert \partial _{k_{x}}\omega _{+-}\right\vert }\right\vert
_{k_{x}=k_{\pm }}=\frac{B^{4}J_{\parallel }^{4}}{8\left( \frac{\omega ^{2}}{4%
}\right) ^{3}\sqrt{J^{2}\omega ^{2}-4B^{2}J_{\parallel }^{2}}}.
\end{equation}

Then, we obtain the jerk current%
\begin{equation}
\frac{\sigma _{\text{jerk}}^{x;x^{3}}}{\sigma _{\text{jerk}}^{\left(
3\right) }}\equiv \frac{\sigma _{\text{inject}}^{x;x^{3}}}{\sigma _{\text{%
inject}}^{\left( 3\right) }}=\theta \left( \omega -\omega _{\text{cr}%
}\right) \frac{64B^{4}J_{\parallel }^{4}}{\omega ^{6}\sqrt{J^{2}\omega
^{2}-4B^{2}J_{\parallel }^{2}}}.
\end{equation}%
It diverges at the optical band edge $\omega =\omega _{\text{cr}}$. When $%
J_{z}=0$,%
\begin{equation}
\frac{\sigma _{\text{jerk}}^{x;x^{3}}}{\sigma _{\text{jerk}}^{0}}=\theta
\left( \omega -2\left\vert B\right\vert \right) \frac{64B^{4}J_{\parallel
}^{3}}{\omega ^{6}\sqrt{\omega ^{2}-4B^{2}}}.
\end{equation}%
On the other hand, when $J_{\parallel }=0$, we have $\sigma _{\text{jerk}%
}^{x;x^{3}}=0$.

\subsubsection{crackle current}

The crackle current is similarly calculated as%
\begin{align}
\frac{\sigma _{\text{crackle}}^{x;x^{5}}}{\sigma _{\text{crackle}}^{\left(
5\right) }}& \equiv \frac{\sigma _{\text{inject}}^{x;x^{5}}}{\sigma _{\text{%
inject}}^{\left( 5\right) }}  \notag \\
& =\theta \left( \omega -\omega _{\text{cr}}\right) \frac{%
3072B^{4}J_{\parallel }^{4}\left( J^{2}\omega ^{2}-5B^{2}J_{\parallel
}^{2}\right) }{\omega ^{10}\sqrt{J^{2}\omega ^{2}-4B^{2}J_{\parallel }^{2}}}.
\end{align}

\subsubsection{higher-order injection current}

The even-order injection current is zero when $J_{z}=0$ because there is
inversion symmetry in the system. On the other hand, it is zero when $%
J_{\parallel }=0$ because the quantum metric is proportional to $%
J_{\parallel }$ as in Eq.(\ref{HeCone}). Explicit calculations show that
they are zero even for $J_{z}\neq 0$ and $J_{\parallel }\neq 0$.
Higher-order injection current is calculated in a similar manner. The $%
\omega $ dependence of log$_{10}\left\vert \sigma _{\text{inject}%
}^{x;x^{\ell }}/\sigma _{\text{inject}}^{\left( \ell \right) }\right\vert $
is shown in Fig.\ref{FigHighPhoto}(a). The asymptotic behavior for large $%
\omega $ is given by%
\begin{equation}
\frac{\sigma _{\text{Inject}}^{x;x^{2\ell ^{\prime }+1}}}{\sigma _{\text{%
Inject}}^{\left( ^{2\ell ^{\prime }+1}\right) }}\propto \theta \left( \omega
-\omega _{\text{cr}}\right) \frac{B^{4}J_{\parallel }^{2}J^{2\ell ^{\prime
}-2}}{\omega ^{2\ell ^{\prime }+5}}.
\end{equation}%
for $\ell ^{\prime }\geq 1$.

\subsection{Higher-order shift current}

The even-order shift current is zero when $J_{z}=0$ because there is
inversion symmetry in the system. On the other hand, it is zero when $%
J_{\parallel }=0$ because the quantum connection is proportional to $%
J_{\parallel }$ as in Eq.(\ref{gxx}). Explicit calculations show that they
are zero even for $J_{z}\neq 0$ and $J_{\parallel }\neq 0$.

Higher-order shift currents are calculated.%
\begin{align}
\frac{\sigma _{\text{shift}}^{x;x^{3}}}{\sigma _{\text{shift}}^{\left(
3\right) }}& =\theta \left( \omega -\omega _{\text{cr}}\right) \frac{%
64B^{2}J_{\parallel }^{2}\sqrt{J^{2}\omega ^{2}-4B^{2}J_{\parallel }^{2}}}{%
\omega ^{8}}, \\
\frac{\sigma _{\text{shift}}^{x;x^{5}}}{\sigma _{\text{shift}}^{\left(
5\right) }}& =\theta \left( \omega -\omega _{\text{cr}}\right) \frac{%
3072B^{2}J_{\parallel }^{2}\sqrt{J^{2}\omega ^{2}-4B^{2}J_{\parallel }^{2}}}{%
\omega ^{13}}  \notag \\
& \times \left( J^{2}\omega ^{2}-8B^{2}J_{\parallel }^{2}\right) .
\end{align}%
The $\omega $ dependence of log$_{10}\left\vert \sigma _{\text{shift}%
}^{x;x^{\ell }}/\sigma _{\text{shift}}^{\left( \ell \right) }\right\vert $
is shown in Fig.\ref{FigHighPhoto}(b). The asymptotic behavior for large $%
\omega $ is given by%
\begin{equation}
\frac{\sigma _{\text{shift}}^{x;x^{2\ell ^{\prime }+1}}}{\sigma _{\text{shift%
}}^{0}}\propto \theta \left( \omega -\omega _{\text{cr}}\right) \frac{%
B^{2}J_{\parallel }^{2}J^{2\ell ^{\prime }-1}}{\omega ^{4\ell ^{\prime }+3}}
\end{equation}%
for $\ell ^{\prime }\geq 1$.

\subsection{Finite temperature effects}

Finally, we study effects on finite temperature. The jerk current at finite
temperature is shown as a function of $\omega $ in Fig.\ref{FigJerkTemp}(a).
The jerk current decreases for high temperature. The third-order shift
current at finite temperature is shown as a function of $\omega $ in Fig.\ref%
{FigJerkTemp}(b). It also decreases for high temperature.

They are understood as follows. The Fermi distribution function difference
is given by 
\begin{equation}
f_{-+}=\frac{1}{\left( \exp \frac{\frac{\hbar ^{2}k_{x}^{2}}{2m}+\frac{%
\omega }{2}-\mu }{k_{\text{B}}T}+1\right) }-\frac{1}{\left( \exp \frac{\frac{%
\hbar ^{2}k_{x}^{2}}{2m}-\frac{\omega }{2}-\mu }{k_{\text{B}}T}+1\right) }.
\end{equation}%
We plot $f_{-+}$ as a function of $\omega $ in Fig.\ref{FigTemp}(a). It
significantly decreases at the optical band edge. We also plot $f_{-+}$ at
the optical band edge in Fig.\ref{FigTemp}(b). $f_{-+}$ is analytically
given by%
\begin{equation}
f_{-+}=\tanh \frac{\left\vert B\right\vert }{2k_{\text{B}}T}
\end{equation}%
at the optical band edge $\omega =2\left\vert B\right\vert $, where we used $%
k=0$ and $\omega =2\left\vert B\right\vert $. It monotonically decreases as
a function of $T$. The asymptotic behavior is analytically given as%
\begin{equation}
\lim_{\omega \rightarrow \infty }f_{-+}=\exp \left[ \frac{-\omega ^{2}}{%
8J^{2}mk_{\text{B}}T}\right]
\end{equation}%
for large $\omega $.

\section{Conclusion}

We have generalized the second-order photovoltaic currents to $\ell $-th
order photovoltaic currents with an arbitrary $\ell $. The $\ell $th-order
injection (shift) current dominates the $\ell $th-order shift (injection)
current in the clean (dirty) sample limit $\tau \rightarrow \infty $ ($\tau
\rightarrow 0$). Then, we have calculated them in $p$-wave magnets and
obtained various analytic results. These results will be useful for future
experiments.

The author is grateful to M. Hirschberger and S. Okumura for helpful
discussions on the subject. This work is supported by CREST, JST (Grants No.
JPMJCR20T2) and Grants-in-Aid for Scientific Research from MEXT KAKENHI
(Grant No. 23H00171).


\begin{thebibliography}{99}
\bibitem{Beli} V. I. Belinicher and B. I. Sturman, The photogalvanic effect
in media lacking a center of symmetry, Sov. Phys. Usp. 23, 199 (1980).

\bibitem{Kraut} W. Kraut and R. von Baltz, Anomalous bulk photo-voltaic
effect in ferroelectrics: A quadratic response theory, Phys. Rev. B 19, 1548
(1979).

\bibitem{Baltz} V. Baltz, A. Manchon, M. Tsoi, T. Moriyama, T. Ono, and Y.
Tserkovnyak, Antiferromagnetic spintronics, Rev. Mod. Phys. 90, 015005
(2018).

\bibitem{Sipe} J. E. Sipe and A. I. Shkrebtii, Second-order optical response
in semiconductors, Phys. Rev. B 61, 5337 (2000).

\bibitem{Frid} V. M. Fridkin, Bulk photovoltaic effect in noncentrosymmetric
crystals, Crystallogr. Rep. 46, 654 (2001).

\bibitem{Ave} C. Aversa and J. E. Sipe, Nonlinear Optical Susceptibilities
of Semiconductors: Results with a Length-Gauge Analysis, Phys. Rev. B 52,
14636 (1995)

\bibitem{AhnX} J. Ahn, G.-Y. Guo, N. Nagaosa, Low-Frequency Divergence and
Quantum Geometry of the Bulk Photovoltaic Effect in Topological Semimetals,
Phys. Rev. X 10, 041041 (2020).

\bibitem{AhnNP} J. Ahn, G.-Y. Guo, N. Nagaosa, A. Vishwanath, Riemannian
geometry of resonant optical responses, Nature Physics 18, 290 (2022)

\bibitem{Juan} F. de Juan, Y. Zhang, T. Morimoto, Y. Sun, J. E. Moore, and
A.G. Grushin, Difference Frequency Generation in Topological Semimetals,
Phys. Rev. Research 2, 012017 (2020).

\bibitem{JuanNC} Fernando de Juan, Adolfo G. Grushin, Takahiro Morimoto,
Joel E. Moore, Quantized circular photogalvanic effect in Weyl semimetals,
Nature Communications 8, 15995 (2017).

\bibitem{Okumura} Shun Okumura, Takahiro Morimoto, Yasuyuki Kato, and
Yukitoshi Motome, Quadratic optical responses in a chiral magnet, Phys. Rev.
B 104, L180407 (2021)

\bibitem{WatanabeInject} Hikaru Watanabe and Youichi Yanase, Chiral
Photocurrent in Parity-Violating Magnet and Enhanced Response in Topological
Antiferromagnet, Phys. Rev. X, 11, 011001 (2021)

\bibitem{Dai} Z. Dai and A. M. Rappe, Recent progress in the theory of bulk
photovoltaic effect, Chemical Physics Reviews 4, 011303 (2023).

\bibitem{EzawaVolta} M. Ezawa, Bulk photovoltaic effects in altermagnets,
Phys. Rev. B 111, L201405 (2025)

\bibitem{Young} S. M. Young and A. M. Rappe, First Principles Calculation of
the Shift Current Photovoltaic Effect in Ferroelectrics, Phys. Rev. Lett.
109, 116601 (2012).

\bibitem{Young2} S. M. Young, F. Zheng, and A. M. Rappe, First-Principles
Calculation of the Bulk Photovoltaic Effect in Bismuth Ferrite, Phys. Rev.
Lett. 109, 236601 (2012).

\bibitem{MorimotoScAd} T. Morimoto and N. Nagaosa, Topological nature of
nonlinear optical effects in solids, Science Advances 2, e1501524 (2016)

\bibitem{Kim} Kun Woo Kim, Takahiro Morimoto, and Naoto Nagaosa, Shift
charge and spin photocurrents in Dirac surface states of topological
insulator, Phys. Rev. B, 95, 035134 (2017)

\bibitem{Barik} T. Barik and J. D. Sau, Nonequilibrium nature of nonlinear
optical response: Application to the bulk photovoltaic effect, Phys. Rev. B
101, 045201 (2020).

\bibitem{Yoshida} Hiroki Yoshida and Shuichi Murakami, Diverging shift
current responses in the gapless limit of two-dimensional systems, Phys.
Rev. B 111, 155402 (2025)

\bibitem{Ma2} Junchao Ma, Qiangqiang Gu, Yinan Liu, Jiawei Lai, Peng Yu,
Xiao Zhuo, Zheng Liu, Jian-Hao Chen, Ji Feng and Dong Sun,Nonlinear
photoresponse of type-II Weyl semimetals, Nature Materials 18, 476 (2019)

\bibitem{Jerk} B. M. Fregoso, R. A. Muniz and J.E. Sipe, Jerk Current: A
Novel Bulk Photovoltaic Effect, Phys. Rev. Lett. 121, 176604 (2018)

\bibitem{JerkComment} G. B. Ventura, D.J. Passos, J. M. Viana Parente Lopes,
and J. M. B. Lopes dos Santos, Comment on "Jerk Current: A Novel Bulk
Photovoltaic Effect" Phys. Rev. Lett. 126, 259701 (2021)

\bibitem{JerkReply} B. M. Fregoso, R. A. Muniz and J.E. Sipe, "Fregoso,
Muniz, and Sipe Reply" Phys. Rev. Lett. 126, 259702 (2021)

\bibitem{Park} Daniel E. Parker, Takahiro Morimoto, Joseph Orenstein, and
Joel E. Moor, Diagrammatic approach to nonlinear optical response with
application to Weyl semimetals, Phys. Rev. B 99, 045121 (2019)

\bibitem{snap} Benjamin M. Fregoso, Bulk photovoltaic effects in the
presence of a static electric field, Phys. Rev. B 100, 064301 (2019):
Erratum, Phys. Rev. B \ 102, 059901(E) (2020)

\bibitem{Provost} J. P. Provost and G. Vallee, Riemannian structure on
manifolds of quantum states, Comm. Math. Phys. 76, 289 (1980).

\bibitem{Ma} Yu-Quan Ma, Shu Chen, Heng Fan, and Wu-Ming Liu, Abelian and
non-Abelian quantum geometric tensor, Phys. Rev. B 81, 245129 (2010).

\bibitem{Holder} T. Holder, D. Kaplan, B. Yan, Consequences of
time-reversal-symmetry breaking in the light-matter interaction: Berry
curvature, quantum metric, and diabatic motion, Phys. Rev. Res. 2, 033100
(2020).

\bibitem{Bhalla} P. Bhalla, K. Das, D. Culcer, A. Agarwal, Resonant
Second-Harmonic Generation as a Probe of Quantum Geometry, Phys. Rev. Lett.
129, 227401 (2022)

\bibitem{Souza} Ivo Souza, Tim Wilkens, and Richard M. Martin, Polarization
and localization in insulators: Generating function approach, Phys. Rev. B
62, 1666 (2000)

\bibitem{WChen2022} W. Chen and G. von Gersdorff, Measurement of
interaction-dressed Berry curvature and quantum metric in solids by optical
absorption, SciPost Phys. Core 5, 040 (2022).

\bibitem{Sousa} Matheus S. M. de Sousa, Antonio L. Cruz, and Wei Chen,
Mapping quantum geometry and quantum phase transitions to real space by a
fidelity marker, Phys. Rev. B 107, 205133 (2023)

\bibitem{Ghosh} Barun Ghosh, Yugo Onishi, Su-Yang Xu, Hsin Lin, Liang Fu and
Arun Bansil, Probing quantum geometry through optical conductivity and
magnetic circular dichroism, Science Advances: sciadv.ado1761 (2024)

\bibitem{Onishi} Yugo Onishi and Liang Fu, Fundamental Bound on Topological
Gap, Phys. Rev. X 14, 011052 (2024)

\bibitem{WChen2024} Wei Chen, Quantum geometrical properties of topological
materials, J. Phys.: Condens. Matter 37 025605 (2025)

\bibitem{WChen2025} Wei Chen, Dielectric and optical markers originating
from quantum geometry, Phys. Rev. B 111, 085202 (2025)

\bibitem{EzawaQG} M. Ezawa, Analytic approach to quantum metric and optical
conductivity in Dirac models with parabolic mass in arbitrary dimensions,
Phys. Rev. B 110, 195437 (2024)

\bibitem{Oh} Chang-geun Oh, Sun-Woo Kim, Kun Woo Kim, Bartomeu Monserrat,
and Jun-Won Rhim, Universal Optical Conductivity from Quantum Geometry in
Quadratic Band-Touching Semimetals, arXiv:2503.18372

\bibitem{JiangRev} Yiyang Jiang, Tobias Holder, and Binghai Yan, Revealing
Quantum Geometry in Nonlinear Quantum Materials, arXiv:2503.04943.

\bibitem{pwave} Anna Birk Hellenes, Tomas Jungwirth, Jairo Sinova, Libor 
\v{S}mejkal, Unconventional p-wave magnets, arXiv:2309.01607.

\bibitem{Martin} I. Martin and A. F. Morpurgo, Majorana fermions in
superconducting helical magnets, Phys. Rev. B 85, 144505 (2012).

\bibitem{H20} S. Hayami, Y. Yanagi, and H. Kusunose, Spontaneous
antisymmetric spin splitting in noncollinear antiferromagnets without
spin-orbit coupling, Phys. Rev. B 101, 220403(R) (2020).

\bibitem{H20b} S. Hayami, Y. Yanagi, and H. Kusunose, Bottom-up design of
spin-split and reshaped electronic band structures in spin-orbit-coupling
free antiferromagnets: Procedure on the basis of augmented multipoles, Phys.
Rev. B 102, 144441 (2020).

\bibitem{H22} S. Hayami, Mechanism of antisymmetric spin polarization in
centrosymmetric multiple-Q magnets based on bilinear and biquadratic spin
cross products, Phys. Rev. B 105, 024413 (2022).

\bibitem{Kuda} Y.B. Kudasov, Topological band structure due to modified
kramers degeneracy for electrons in a helical magnetic field, Phys. Rev. B
109, L140402 (2024).

\bibitem{EzawaPwave} M. Ezawa, Topological insulators based on $p$-wave
altermagnets; Electrical control and detection of the altermagnetic domain
wall, Phys. Rev. B 110, 165429 (2024).

\bibitem{Edel} M. Ezawa, Out-of-plane Edelstein effects: Electric-field
induced magnetization in p-wave magnets, Physical Review B 111 (16), L161301
(2025)

\bibitem{EzawaPNeel} M. Ezawa, Purely electrical detection of the Neel
vector of p-wave magnets based on linear and nonlinear conductivities,
arXiv:2410.21854

\bibitem{Yamada} Rinsuke Yamada, Max T. Birch, Priya R. Baral, Shun Okumura,
Ryota Nakano, Shang Gao, Yuki Ishihara, Kamil K. Kolincio, Ilya Belopolski,
Hajime Sagayama, Hironori Nakao, Kazuki Ohishi, Taro Nakajima, Yoshinori
Tokura, Taka-hisa Arima, Yukitoshi Motome, Moritz M. Hirschmann, and Max
Hirschberger, Gapping the spin-nodal planes of an anisotropic p-wave magnet
to induce a large anomalous Hall effect, arXiv:2502.10386

\bibitem{Comin} Qian Song, Srdjan Stavric, Paolo Barone, Andrea Droghetti,
Daniil S. Antonenko, Jorn W. F. Venderbos, Connor A. Occhialini, Batyr
Ilyas, Emre Ergecen, Nuh Gedik, Sang-Wook Cheong, Rafael M. Fernandes,
Silvia Picozzi, and Riccardo Comin, Electrical switching of a p-wave magnet,
Nature 642, 64 (2025)

\bibitem{Elliptic} M. Ezawa, Quantum geometry and elliptic optical dichroism
in $p$-wave magnets Motohiko Ezawa, Phys. Rev. B 112, 045302 (2025)

\bibitem{Maeda} Kazuki Maeda, Bo Lu, Keiji Yada, Yukio Tanaka, Theory of
tunneling spectroscopy in p-wave altermagnet-superconductor hybrid
structures, J. Phys. Soc. Jpn. 93, 114703 (2024).

\bibitem{Fukaya} Yuri Fukaya, Bo Lu, Keiji Yada, Yukio Tanaka, Jorge Cayao,
Superconducting phenomena in systems with unconventional magnets,
arXiv:2502.15400

\bibitem{Matt} Matt Visser, Jerk, snap and the cosmological equation of
state, Class. Quantum Grav. 21 2603 (2004)

\bibitem{Brek} Bjonulf Brekke, Pavlo Sukhachov, Hans Glokner Giil, Arne
Brataas, Jacob Linder, Minimal models and transport properties of
unconventional p-wave magnets, Phys. Rev. Lett. 133, 236703 (2024)

\bibitem{GI} M. Ezawa, Third-order and fifth-order nonlinear spin-current
generation in g-wave and i-wave altermagnets and perfectly nonreciprocal
spin current in f-wave magnets, Phys. Rev. B 111, 125420\ (2025)

\bibitem{Matsuura} Shunji Matsuura and Shinsei Ryu, Momentum space metric,
nonlocal operator, and topological insulators, Phys. Rev. B 82, 245113 (2010)

\bibitem{Gers} G. von Gersdorff and W. Chen, Measurement of topological
order based on metric-curvature correspondence, Phys. Rev. B 104, 195133
(2021).
\end{thebibliography}
\end{document}